\documentclass[twocolumn]{aastex631}
\usepackage{amsmath}

\newcommand{\eb}{\begin{equation}}
\newcommand{\ee}{\end{equation}}
\newcommand{\bp}{$G_{\rm BP}$}
\newcommand{\rp}{$G_{\rm RP}$}

\usepackage{subfigure,graphicx}
\usepackage{url}
\usepackage{color}
\definecolor{darkgray}{gray}{0.4}
\definecolor{patinared}{rgb}{.72,.10,0}
\definecolor{patinablue}{rgb}{0,.20,.65} 
\definecolor{orange}{rgb}{1,0.5,0}
\definecolor{rkka}{RGB}{219,66,32}

\hypersetup{
  colorlinks=True,      
  urlcolor=darkgray,    
  linkcolor=patinared,  
  citecolor=patinablue, 
}

\shorttitle{Variability of Gaia CRF sources}
\shortauthors{Makarov}
\graphicspath{{./}{figures/}}

\begin{document}

\title{Optical variability of Gaia CRF3 sources with robust statistics\\
and the 5000 most variable quasars}

\correspondingauthor{Valeri Makarov}
\email{valeri.v.makarov.civ@us.navy.mil}

\author[0000-0003-2336-7887]{Valeri V.\ Makarov}
\affiliation{U.S. Naval Observatory, 3450 Massachusetts Ave NW, Washington, DC 20392-5420, USA}



\begin{abstract}

Using the light curve time series data for more than 11.7 million variable sources published in the Gaia Data Release 3, the average magnitudes, colors, and variability parameters have been computed for 0.836 million Gaia CRF objects, which are mostly quasars and active galactic nuclei (AGNs). To mitigate the effects of occasional flukes in the data, robust statistical measures have been employed, namely, the median, median absolute deviation, and Spearman correlation. We find that the majority of the CRF sources have moderate amplitudes of variability in the Gaia $G$ band just below 0.1 mag. The heavy-tailed distribution of variability amplitudes (quantified as robust standard deviations) does not find a single analytical form, but is closer to Maxwell distribution with a scale of 0.078 mag. The majority of CRF sources have positive correlations between $G$ magnitude and $G_{\rm BP}-G_{\rm RP}$ colors, meaning that these quasars and AGNs become bluer when they are brighter. The variations in the \bp\ and \rp\ bands are also mostly positively correlated. Dependencies of all variability parameters with cosmological redshift are fairly flat for the more accurate estimates above redshift 0.7, while the median color shows strong systematic variations with redshift. Using a robust normalized score of magnitude deviations, a sample of 5000 most variable quasars is selected and published. The intersection of this sample with the ICRF3 catalog shows a much higher rate of strongly variable quasars (mostly, blazars) in ICRF3. 

\end{abstract}


\section{Introduction} \label{int.sec}

The scrupulously constructed collection of QSO-like objects (candidate quasars and active galactic nuclei, or AGNs) called The Gaia Celestial Reference Frame (CRF) catalog, serves two primary purposes \citep{2018A&A...616A..14G, 2022A&A...667A.148G}. First, an intrinsic rank-6 degeneracy of the Gaia global astrometric solution is removed. A relatively small intersection of it with the optical counterparts of the International Celestial Reference Frame catalog \citep[ICRF3, ][]{2020A&A...644A.159C} is used to align the geometric orientation of the two Cartesian coordinate systems, which is a matter of practical convenience for numerous applications of the fundamental celestial frame. The intrinsically indeterminate spin of the entire proper motion system is adjusted to posterior net zero using a much greater sample of Gaia CRF sources. This adjustment has deep cosmological and even philosophical implications, because it concerns the motion of the universe as a whole \citep{2010IAUS..261..345M}. On the technical side, reliable CRF sources can also be internally used to calibrate systematic astrometry effects and validate the results.

The accuracy and long-term reliability of these applications are inherently related to the photometric properties of CRF sources. Quasars (irrespective of their brightness at radio wavelengths) and AGNs are known to be variable sources of electromagnetic radiation across the entire spectrum \citep{1997ARA&A..35..445U}. The time scale of their variations ranges from intra-night to years and decades. The morphology of optical light curves is complex and mostly stochastic, or unpredictable. Even the physical mechanism of variability and its origin is not a settled issue, with presented hypotheses including the relativistic jets, massive accretion disks in the central engines, and external gravitational microlensing. The impact of optical variability of CRF objects on the accuracy of astrometric applications and, generally, the quality of the fundamental reference frame is significant.

Epoch photometry is an important integral part of the Gaia mission Data Release 3 \citep[DR3, ][]{2016A&A...595A...1G, 2021A&A...649A...1G, 2023A&A...674A...1G}. At the level of pixel photon count processing, the estimation of astrometric photocenters and photometric amplitudes are interdependent when the latter are derived from the image fitting. This is the case with the Gaia photometry in the broad ``whitelight" $G$ band based on the data collected with the astrometric CCDs, which involves PSF or LSF fitting \citep{2018A&A...616A...3R}. On the other hand, the \bp\ and \rp\ magnitudes\footnote{Throughout this text, \bp\ and \rp\ magnitudes are intermittently abbreviated to BP and RP magnitudes to avoid double subscripts} corresponding to the narrower BP and RP bands, covering 330--680 nm and 630--1050 nm ranges, respectively, were derived in a different way, closer to the aperture photometry method. The low-resolution spectra on the dedicated spectroscopic CCDs were integrated within a window of 60 samples. The derived BP and RP fluxes are more sensitive to the background and stray light calibration, and acquire additional complications from the different gating schemes. BP and RP measurements become problematic for unresolved double sources, extended images (galaxies and nebulae), and crowded areas. Such impairments are adequately captured by the {\tt phot\_bp\_rp\_excess\_factor} parameter in the main Gaia DR3 catalog, which reflects the degree of coherence between the mean $G$, BP, and RP determinations for each source. In application to quasars and AGNs, this parameter is strongly perturbed for relatively nearby objects with redshifts below 0.5 \citep{2022ApJ...933...28M}. The principles of photometric data reductions and DR3 photometry validation are discussed in \citep{2021A&A...649A...3R}.

The main objective of this paper is to quantify the general variability properties of a global sample of quasars and AGNs, used for the alignment of Gaia CRF, on an unprecedented scale by using the 0.8 million objects cross-matched in Gaia DR3 CRF and epoch photometry catalogs. The pre-Gaia efforts to estimate the variability amplitudes of the most important reference frame objects included only dozens of sources and required coordinated long-term observational campaigns on small and medium-sized telescopes \citep{2013A&A...552A..98T, 2016A&A...587A.112T}. A larger dataset of multi-band epoch photometry from the Pan-STARRS survey was used by \citet{2021AJ....162...21B}, but their study was limited to the optical sample of ICRF3. Much larger samples of spectroscopically confirmed quasars emerged from the Sloan Digital Sky Survey \citep[SDSS, ][]{2000AJ....120.1579Y}, providing impetus to the studies of their basic photometric properties \citep[e.g., ][]{2004ApJ...601..692V, 2011A&A...525A..37M}, but the poor cadence forced the authors to use ensemble (i.e., sample-average) statistical estimation. In this paper, {\it en masse} estimation of individual variability properties is performed for an order of magnitude larger collection of sources.

The most variable sources can also be used in the novel method of detection of dual AGNs and quasars via the Variability-Induced Motion \citep[VIM, ][]{2022ApJ...933...28M} or varstrometry \citep{2020ApJ...888...73H} effects, where the uncorrelated brightness variations of the two close nuclei produce a measurable astrometric shift of the unresolved photocenter  along the line connecting them. The effect is very prominent for double stars in the simultaneous {\it Kepler} mission astrometric and photometric data \citep{2016ApJS..224...19M}. The relevance of this study for fundamental astrometry is also seen in the currently active investigation of the position offsets between the radio celestial frame sources and their optical counterparts in Gaia \citep{2017ApJ...835L..30M, 2017MNRAS.471.3775P, 2019MNRAS.482.3023P, 2019ApJ...873..132M}. It concerns at least a quarter of the common sources, which weakens the link between the two celestial reference frames. The main hypothesis is that radio core shifts in the relativistic jets are responsible \citep{2021A&A...651A..64L}, but optical photocenter displacements cannot be ruled out either, in which case correlations with optical variability may be present. Anticorrelation between the magnitude of radio-optical offsets and the degree of optical  variability \citep{2022ApJ...939L..32S}, which is consistent with the radio core shift hypothesis, implies that the highly variable blazars are the best sources from the astrometric point of view. 

Although the derived characteristics of photometrically variable AGNs have been collected in the GLEAN catalog \citep{2023A&A...674A..24C}, we perform here a new analysis from first principles focusing on the much larger Gaia CRF3 sample, and employing basic principles of robust statistical analysis. This allows us to collect the light curves for the most interesting objects and investigate additional correlations and trends that have not been considered in previous publications. 

\section{Initial samples and data}
Making use of the online TAP VizieR facility, the massive data table with epoch photometry measures (556 million records)\footnote{\url{http://vizier.cds.unistra.fr/viz-bin/VizieR-3?-source=I/355/epphot}} was matched with the Gaia DR3 CRF table\footnote{\url{http://vizier.cds.unistra.fr/viz-bin/VizieR-3?-source=I/355/gcrf3xm}}, which lists 1.6 million sources, by the common Gaia source identifiers {\tt source\_id}. The resulting data set includes over 35 million individual epoch photometry measurements of 835,291 CRF objects with the broadband $G$ magnitudes available after filtering out entries with the {\tt noisyFlag} and {\tt GrVFlag} values set to 1. These flags indicate unreliable and possibly corrupted observations, which occurred due to registered events of instrument malfunction, such as decontamination runs,\footnote{see \url{https://gea.esac.esa.int/archive/documentation/GDR3/Introduction/chap_cu0int/cu0int_sec_release_framework/cu0int_sssec_spacecraft_status.html} for details} or rejected by the variability processing pipeline for other reasons. The applied filtering helps to avoid propagation of false positives to the sample of variable sources, as a significant fraction of epoch photometry values is affected by calibration glitches, special instrumental events, and statistical flukes.  

In addition to the quality flags filtering, the general sample of CRF sources (hereafter, the CRF-sample) considered in this paper includes only objects with more than 10  single-epoch measurements. Marginally faint or transient sources with few legitimate measurements have been excluded. Smaller subsets of this sample were separately analyzed: the $BR$-sample of 816,686 objects, which also have more than 10 accepted measurements in each of the \bp\ and \rp\ filters (after filtering out the corresponding quality flags {\tt BPrFlag} and {\tt RPrFlag}), and the ICRF-sample of 2690 objects, which have definitive cross-identification names in the {\tt IERSname} field of the CRF table. The latter subset of the $BR$-sample includes the optical counterparts of the ICRF3 catalog with the filtered epoch photometry data from Gaia DR3. The smallest number of single-epoch $G$ magnitudes per object is (by construction) 11, the median is 36, and the maximum number is 234. From the general CRF-sample, using a uniform selection criterion, a subset of 5033 most variable objects (MVO-sample) is constructed and published as a catalog (Section \ref{mvo.sec}). Individual light curves for these MVO sources (mostly, blazars) have been retained and visually reviewed. Finally, 0.3 million brighter sources from the CRF-sample with median $G$ magnitudes brighter than 19.6 were cross-matched with the Quaia catalog of spectroscopic/photometric redshifts to produce a $z$-sample of  quasars with uniformly estimated redshifts counting 298,487 sources.

\section{Robust photometric parameters}  \label{ex.sec}
The formal errors of fluxes in the Gaia epoch photometry table do not capture the actual dispersion of measurements, because quasars and AGNs are intrinsically variable sources of light. Therefore, instead of the traditional mean and weighted mean, standard deviation (or sample rms), and normalized $\chi$ values, we have to employ robust statistical parameters. For each source in the three working samples, I gathered all acceptable measurements  and computed the median $G$ magnitudes:
\eb 
\hat G = {\rm median}\left( G_i\right), \hspace{4mm} i=1,\ldots,n_G,
\ee 
where $n_G$ is the number of single-epoch measurements in $G$. The sample distribution of $\hat G$ magnitudes is peaked at 20.2 and is truncated at 21. To quantify the amplitude of variability, the robust analog of standard deviation is computed for each object as
\eb 
{\rm smad}(\hat G) = 1.5*{\rm median}\left( | \hat G-G_i |\right), \hspace{4mm} i=1,\ldots,n_G,
\label{hatG.eq}
\ee
where smad stands for ``scaled median absolute deviation", and the scaling coefficient 1.5 approximately equals the ratio of the standard deviation to the median absolute deviation (MAD) for the normal distribution. 

Many of the CRF objects are marginally faint, and the robust smad($\hat G$) quantities do not faithfully reflect the physical amplitude of variability, being inflated by the photon noise measurement error. Furthermore, faint quasars are affected by crowding problems at lower Galactic latitudes, where the high density of foreground field stars increase the rate of perturbations at the pixel level. The variable bright sources are generally more reliably identified even if their absolute amplitudes are relatively smaller. A robust analog of the normalized variability score is introduced, which is computed as
\eb 
\hat Z = {\rm median}\left( | \hat G-G_i |*f_i/(1.086*e_i)\right), \hspace{4mm} i=1,\ldots,n_G,
\label{z.eq}
\ee
where $f_i$ is the tabulated measured flux in $G$, and $e_i$ is its formal error. Note that the absolute deviation of $G_i$ from the median value replaces the quadratic deviation from the mean, while the $1.086\,e_i/f_i$ ratio approximates the formal standard deviation of $G_i$. 

In addition to the statistical parameters $\hat G$, ${\rm smad}(\hat G)$, and $\hat Z$, the Spearman's rank correlation (also known as the Spearman's rho) is computed for each of the 0.817 million sources in the $BR$-sample. In difference to the Pearson correlation, the Spearman's correlation is based on the ranking of the two variables, which does not assume a linear dependence between them \citep[][Vol. 2A]{10.5555/59556}. It belongs to the class of order statistics, which are often more robust in the presence of non-Gaussian measurement noise or flukes. Since the residuals of measured magnitudes are explicitly non-Gaussian with respect to the chosen fit (the median, in this case), and the given samples are not linearly dependent, the non-parametric character of the Spearman's correlation provides a crucial advantage. Here we compute the correlation between the tabulated $G_i$ measurements and the $G_{\rm BP}-G_{\rm RP}$ color, which is hereafter denoted as BP$-$RP for simplicity:
$\rho(G,{\rm BP}-{\rm RP})$. This parameter, which can take values between $-1$ and $+1$, quantifies the degree of coherence between the measured deviations of $G$ magnitude and color. A positive Spearman's rho signifies that a positive variation of one parameter is more often associated with a positive variation of the other.

Additional photometric variability parameters were computed for each source in the $BR$-sample, which has sufficient information for the \bp\ and \rp\ bands. These include:
\begin{eqnarray}
    \hat B &=&{\rm median}({\rm BP}_i), \\
    \hat R &=&{\rm median}({\rm RP}_i), \nonumber \\
    \hat C &=&{\rm median}({\rm BP}_i-{\rm RP}_i),\nonumber \\
    {\rm smad}({\rm BP}-{\rm RP}) &=&1.5*{\rm median}\left( | \hat C-{\rm BP}_i+{\rm RP}_i |\right),\nonumber \\
    \rho({\rm BP},{\rm RP}) &=&\rho\left(G_{\rm BP_i},G_{\rm RP_i}\right), \nonumber \\
    &&{\rm for}\hspace{4mm} i=1,\ldots,n_{BR}.\nonumber 
    \label{c.eq}
\end{eqnarray}
The robust analog of standard deviation ${\rm smad}({\rm BP}-{\rm RP})$ represents the amplitude of color variations, while the Spearman's correlation coefficient $\rho({\rm BP},{\rm RP})$ shows the degree of concordance between the deviations of the blue and red magnitudes from their respective median values. A positive value signifies that when the source becomes brighter in one of the bands, it also becomes brighter in the other. A negative correlation coefficient is an interesting occurrence, implying that the general variability is coupled with large changes in the slope of the continuum spectral energy distribution in the optical domain. 

\section{General statistics}

\begin{figure*}
\includegraphics[width=0.45\linewidth]{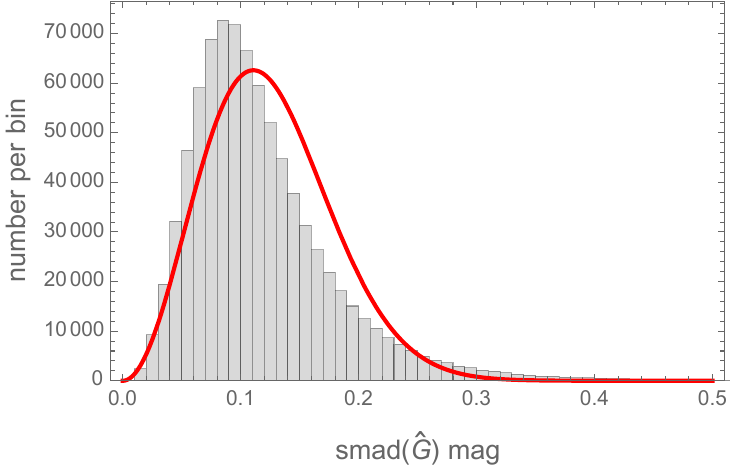}
\includegraphics[width=0.45\linewidth]{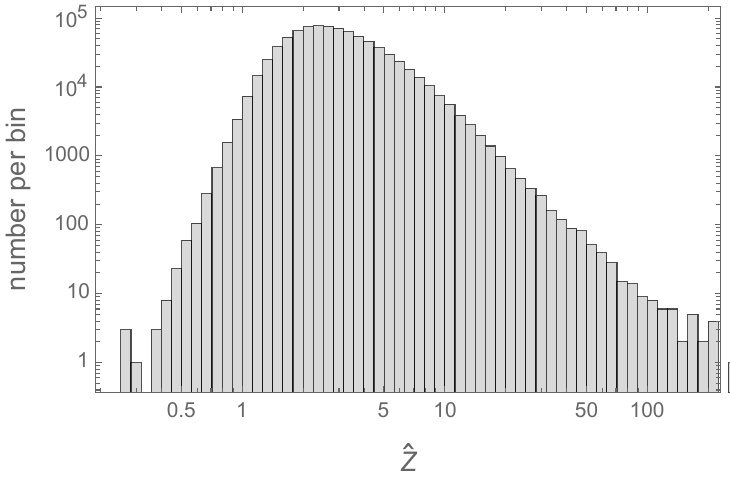}
\caption{Histograms of robust variability parameters in $G$ for the general Gaia CRF sample. Left panel: dispersion of measured magnitudes smad$(\hat G)$. The red curve shows the closest analytical fit, which is Maxwell[0.078] distribution. Right panel: normalized dispersion of measured magnitudes $\hat Z$.}
\label{G.fig}
\end{figure*}

For the general $G$-sample including 0.835 million CRF sources, we find that the robust standard deviation of single-epoch $G$ magnitude measures has an asymmetric distribution peaked at 0.088 mag, with a long tail stretching toward 0.5 mag and beyond (Fig. \ref{G.fig}, left panel). The median of smad$(\hat G)$ is 0.105 mag, 25\% of the values are above 0.145 mag, 10\% are above 0.197 mag, and 1\% is above 0.347 mag. This implies that the bulk of Gaia-selected quasars and AGNs are moderately variable. The histogram of smad$(\hat G)$ does not find a single analytical fit. I have used the {\tt FindDistribution} routine of Wolfram Mathematica with the AIC test criterion, which attempts to fit a great variety of analytical univariate probability density functions with free optimized parameters. The closest approximation, which is shown in the Figure with the red line, is Maxwell[0.078], but its flatter peak is shifted to higher values, and it under-represents the tail above 0.25 mag. To see if this dispersion is significant compared to the measurement error, we refer to the histogram of the normalized score $\hat Z$ in Fig. \ref{G.fig}, right panel (note the logarithmic scale of both axes). The median of
$\hat Z$ values is 2.73, with only 0.7\% of the sample having $\hat Z<1$.

\begin{figure}
\includegraphics[width=0.95\linewidth]{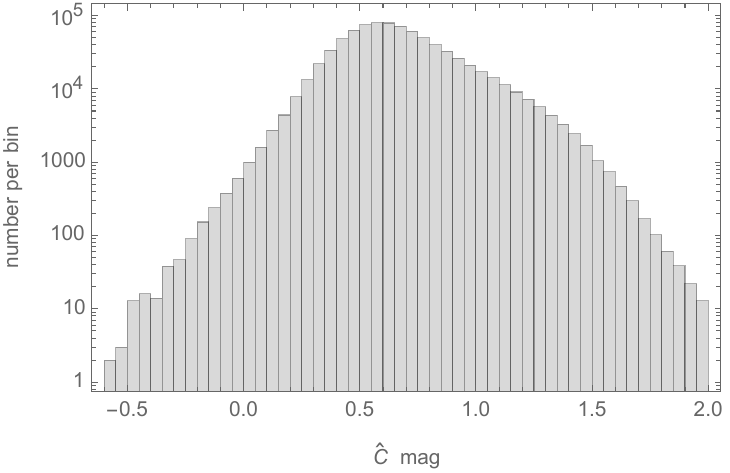}
\caption{Histogram of median \bp$-$\rp\ colors for 0.8 million Gaia CRF3 sources, in logarithmic scale.}
\label{chat.fig}
\end{figure}

The distribution of $\hat C$ colors is peaked at 0.6 mag (not shown for brevity), but there seems to be a vague hint at a secondary population of red quasars with colors around 1.3 mag, see Fig. \ref{chat.fig}. We note, however, that with the distribution of $G$ magnitudes piling up on the faint limit, the statistics related to the \bp\ and \rp\ magnitudes can be affected by the large ``survival" bias \citep{2021A&A...649A...5F}. An intrinsically red and faint source is more likely to trigger a non-detection in the blue band when the collected photon count falls below the detection threshold (1 $e^-$ s$^{-1}$) because of the photon shot noise and detector readout noise fluctuation. The surviving detections increasingly tend to be brighter than the true mean magnitude toward the lower flux limit. The problem is common for magnitude-limited surveys, and in Tycho-1 photometric reductions, a bias-correcting procedure was implemented \citep{1997A&A...325..360H}, which, inevitably, made the estimation more uncertain. Here we conservatively limit our consideration to the bright one-third of the $G$-sample, rejecting all sources fainter than $G=19.6$ mag. This minimizes the detection bias, but the reddest objects may still be affected, resulting in underestimated variability parameters concerning \bp.

\begin{figure*}
\includegraphics[width=0.45\linewidth]{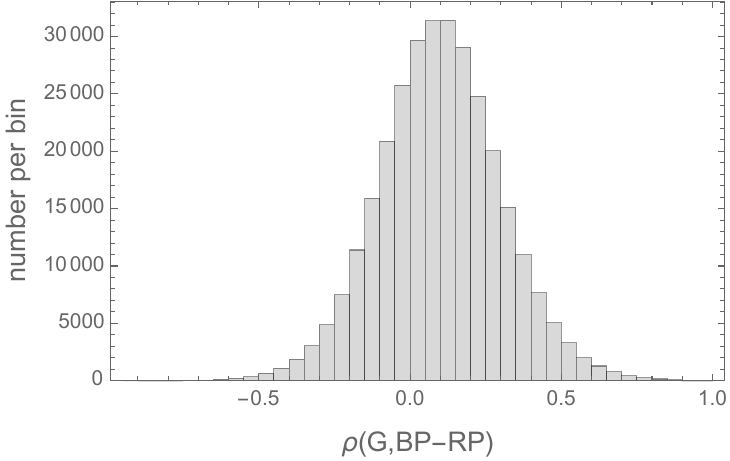}
\includegraphics[width=0.45\linewidth]{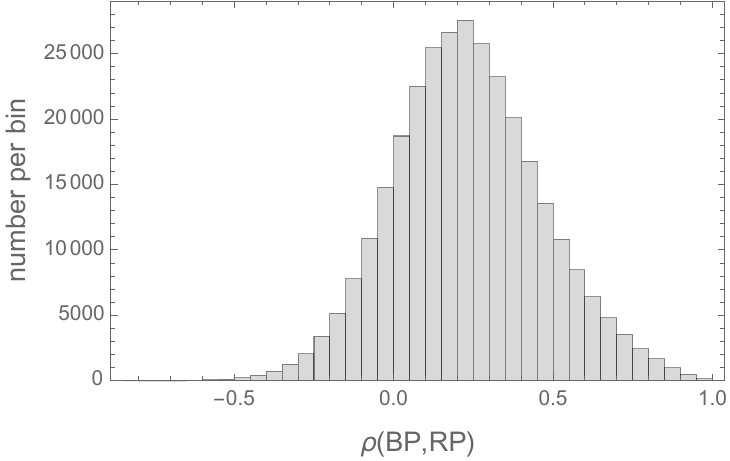}
\caption{Histograms of Spearman's correlation coefficients $\rho(G,{\rm BP}-{\rm RP})$ (left panel) and $\rho({\rm BP},{\rm RP})$ (right panel) for the brighter subset of Gaia CRF3 with $\hat G <19.6$ mag.}
\label{rho.fig}
\end{figure*}

We find two clear trends from the comparison of variability statistics for the all-inclusive $BR$-sample and the bright subsample. The robust smad values are all smaller for the bright sources. For example, the median smad(BP) is reduced from 0.36 for the general sample to 0.22 mag for the bright sample, and smad(RP) drops from 0.32 to 0.19 mag. This can be interpreted as the reduced contribution of random observational errors, i.e., higher accuracy. In agreement with this explanation, the Spearman's correlation coefficients show a counter trend, being considerably larger for the brighter objects. The sample distributions are shown in Fig. \ref{rho.fig} for $\rho(G,{\rm BP}-{\rm RP})$ (left panel) and $\rho({\rm BP},{\rm RP})$ (right panel). In both cases, we find an obvious positive shift from zero. The median value of $\rho(G,{\rm BP}-{\rm RP})$ is 0.098  and the median of $\rho({\rm BP},{\rm RP})$ is 0.224. The conclusion is that CRF objects {\it mostly} become bluer with increasing broadband brightness, and the variations in the blue and the red bands are concordant. The positive correlation between brightness and color is weaker, however.

\begin{figure*}
\includegraphics[width=0.45\linewidth]{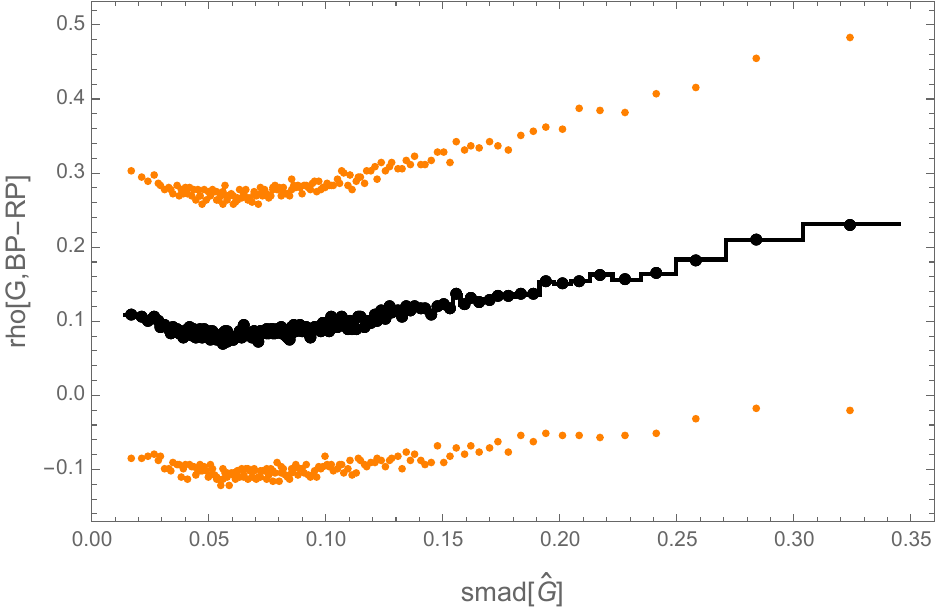}
\includegraphics[width=0.45\linewidth]{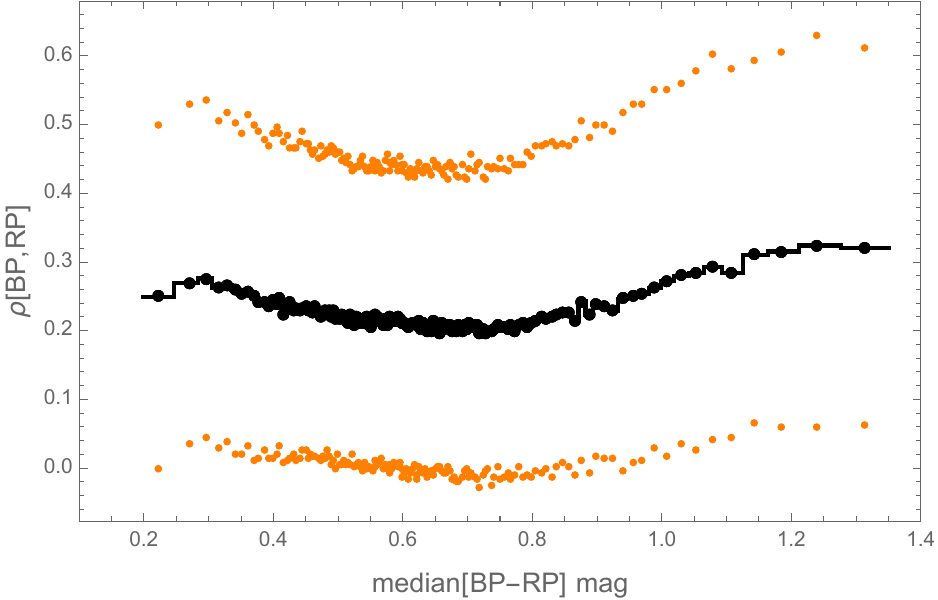}
\caption{Empirical dependence of Spearman's correlation coefficients $\rho(G,{\rm BP}-{\rm RP})$ on variability amplitude smad($\hat G$) (left panel) and $\rho({\rm BP},{\rm RP})$ on the median (BP$-$RP) color (right panel) for the brighter subset of Gaia CRF3 with $\hat G <19.6$ mag. The black dots connected by a step line show the median values for each bin of sorted data. The orange dots show the 0.159 and 0.841 quantiles of the values for each bin, which correspond to the $\pm 1\sigma$ width of the sample distribution.}
\label{bin.fig}
\end{figure*}

To shed light on underlying dependencies bringing up these positive correlations, the bright sample was sorted by various parameters and partitioned into 150 bins of approximately 2050 objects each, and the $\{0.159,0.5,0.841\}$ quantiles of the parameters of interest were computed for each bin. The most conspicuous results for $\rho(G,{\rm BP}-{\rm RP})$ as a function of variability amplitude smad($\hat G$) and $\rho({\rm BP},{\rm RP})$ as a function of median (BP$-$RP) color are shown in Fig. \ref{bin.fig} in left and right panels, respectively. The median abscissa and ordinate values are represented by black dots connected with a step-wise interpolation line. The outer quantile values are shown with orange dots. They correspond to the $\pm1\sigma$ uncertainty interval, or the robust width of the sample distribution. We find that the least variable sources have significantly less positive correlation between $G$ magnitude and color, while the more variable quasars with smad($\hat G$)$>0.06$ mag show a steady trend toward more coherent variations in brightness and color. The concordance of BP and RP variations depends on the median color. The global minimum is achieved at $\hat C\approx$ 0.7 mag, which is close to the modal value for the entire sample. The bluest and the reddest quasars, however, show a greater degree of concordance between BP and RP. In both cases, about 15\% of the sample show a counter-directed correlation with $\rho(G,{\rm BP}-{\rm RP})<-0.1$ and $\rho({\rm BP},{\rm RP})<0$.

\section{The most variable CRF sources} \label{mvo.sec}

Sources with the greatest $G$ magnitude dispersion are not necessarily real highly variable quasars. We have seen that the observational dispersion (i.e., measurements error) becomes a comparable or even dominating contributor at the faint limit of the sample. Furthermore, a significant number of faint sources have only a dozen data points, and the probability that some of them are flukes or gross errors is non-negligible. Such flukes are often associated with specific time periods corresponding to instrumental events, e.g., the decontamination procedures. The following composite parameter was used in this study to reliably select the most variable sources:
\eb 
V=\sqrt{n_G}\;\hat Z,
\ee 
and the sources with $V>100$ have been selected. This criterion gives preference to brighter objects with smaller formal measurement errors, larger amplitudes of variability in $G$, and larger number of accepted observations. With a typical number of $G$ magnitudes of 36, the $\hat Z$-score has to be above $\simeq 17$ for a source to be selected. This corresponds to the farthest part of the sample distribution in Fig. \ref{G.fig}, right. The total number of thus selected most variable objects is 5033.

The light curves of the most variable CRF objects were retained and  visually inspected. This collection may find multiple applications. For example, previously undetected blazars should be present there outside the Sloan Digital Sky Survey (SDSS) footprint. As discussed in Introduction, blazars are preferred objects for replenishing the ICRF sample, if they are found to be sufficiently radio-loud. The visual inspection of the light curves reveals that their morphology can be roughly divided into three overlapping types: 1) noisy or jittery, with large stochastic changes between the consecutive Gaia visits (with a modal separation close to 1 month); 2) relatively smooth and coherent long-term variations over the 2.5-yr DR3 time span; 3) moderately dispersed light curves with pronounced dips or flares lasting for a few months. The Gaia light curves confirm the remarkable diversity of behaviors found in the {\it Kepler} mission data for a much smaller sample of quasars \citep{2018ApJ...857..141S}.

\begin{figure}
\includegraphics[width=0.85\linewidth]{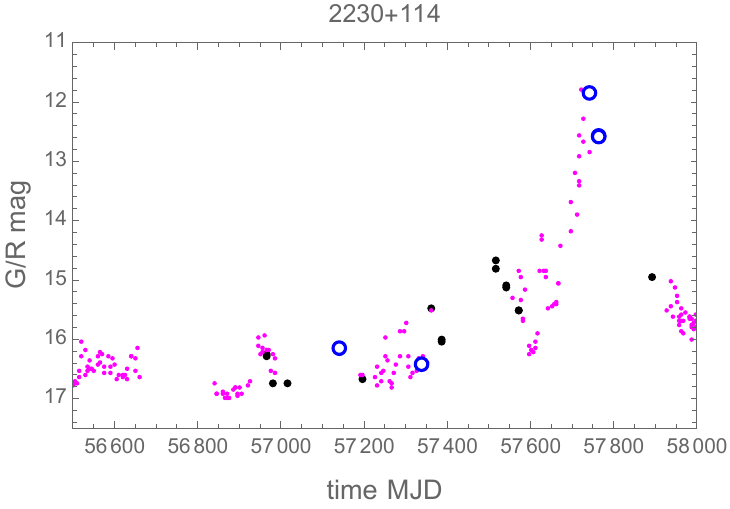}
\caption{Gaia DR3 light curve of the extreme flaring FSRQ blazar QSO J2232+1143 = IERS B2230+114, which is also part of the ICRF3. The filtered measurements that have been accepted in this analysis are shown with black dots. The quality flags in Gaia discard a genuine flare event around MJD = 57750, when the source was brighter than $G=12$ mag. Blue circles represent the discarded Gaia measurements with {\tt noisyFlag}=1. Ground-based photometric measurements from \citep{2019ApJ...880...32L} in the $R$ band are shown with magenta dots.}
\label{22.fig}
\end{figure}

The overlap between this list of most variable quasars and the collection of optical light curves presented by \citet{2019ApJ...880...32L} for 173 Fermi-detected gamma-ray blazars counts 114 objects. For these common objects including both flat-spectrum radio quasars (FSRQ) and BL Lac classes, the ground-based light curves in the (roughly) $R$ band can be visually compared with the Gaia data. One curious example is the extreme flaring FSRQ blazar QSO J2232+1143 = IERS B2230+114, which is normally at $\sim 16$ mag in the optical, but sporadically flares up by more than 4 mag. One such extreme flare was recorded from the ground on MJD = 57705.6 with $R=11.83$ mag. A few transits in Gaia overlap with the time span of this event, and the brightest magnitudes were measured on MJD = 57741.7 at $G=11.85$, BP$=12.22$, and RP$=11.25$. All the measurements corresponding to the flare event, however, were flagged as ``noisy" ({\tt noisyFlag}=1) by the photometric pipeline and discarded. The filtered Gaia light curve in $G$ is shown in Fig. \ref{22.fig} with black dots, and the filtered out measurements with blue circles. The ground-based photometric data in $R$ from \citep{2019ApJ...880...32L} are represented by magenta dots.

\begin{table*}
\caption{Robust variability parameters of 5033 most variable Gaia CRF quasars and AGNs.\label{para.tab}
}
\tiny \hspace*{-2.5cm}
\begin{tabular}{r|rr|cccccccccc}
\hline
Source & RA & Decl. & $\hat G$  & smad($\hat G$) & $\hat Z$ & $n_G$ & $\hat C$ & smad($\hat C$) &smad(BP) &smad(RP) & $\rho(G,$BP$-$RP) & $\rho($BP,RP)\\
\phantom{name} & $\degr$ & $\degr$ & mag & mag & & & mag & mag & mag & mag & & \\
\hline \hline
  3540152563597824 & 46.992156923 & 4.505925179 & 18.616 & 0.399 & 22.331 & 25 & 1.151 & 0.182 & 0.357 & 0.234 & 0.26 & 0.71 \\
 8527129285880704 & 44.862819247 & 7.794345129 & 17.803 & 0.183 & 21.279 & 24 & 1.305 & 0.086 & 0.23 & 0.24 & -0.21 & 0.91 \\
 21647223582925568 & 41.661108028 & 9.765033977 & 18.387 & 0.391 & 21.86 & 22 & 0.821 & 0.185 & 0.437 & 0.292 & 0.59 & 0.62 \\
 68904125970527360 & 52.634617155 & 24.677499431 & 18.457 & 0.373 & 18.616 & 29 & 1.343 & 0.092 & 0.26 & 0.173 & 0.28 & 0.71 \\
 72742078681534976 & 35.390559417 & 11.775383323 & 17.167 & 0.187 & 29.412 & 22 & 0.77 & 0.11 & 0.334 & 0.154 & 0.79 & 0.94 \\
 100986672678118016 & 34.133726641 & 23.247321157 & 18.352 & 0.331 & 23.786 & 23 & 0.879 & 0.144 & 0.168 & 0.297 & 0.34 & 0.85 \\
 105026312757899648 & 29.228427076 & 24.383038148 & 18.979 & 0.671 & 22.174 & 24 & 0.949 & 0.239 & 0.438 & 0.401 & 0.35 & 0.84 \\
 143525957219741952 & 43.770329655 & 39.680872855 & 18.283 & 0.282 & 22.576 & 22 & 0.577 & 0.158 & 0.184 & 0.18 & 0.39 & 0.68 \\
 228323931790519936 & 65.98337388 & 41.834086841 & 19.007 & 0.383 & 18.773 & 49 & 2.123 & 0.433 & 0.32 & 0.25 & 0.04 & 0.53 \\
 268450058890077440 & 85.724753594 & 56.802520356 & 19.047 & 0.415 & 19.151 & 37 & 1.074 & 0.271 & 0.546 & 0.301 & 0.17 & 0.71 \\
 286452229629246720 & 85.030225269 & 62.79807295 & 16.206 & 0.055 & 12.919 & 65 & 0.593 & 0.027 & 0.053 & 0.045 & 0.14 & 0.68 \\
 292822181522982144 & 19.186575609 & 22.554997968 & 18.931 & 0.42 & 17.455 & 33 & 0.674 & 0.332 & 0.567 & 0.29 & 0.55 & 0.62 \\

\hline
\end{tabular}
\newline
Notes: Only the leading 12 rows are reproduced. The full table is available online as a CSV file. Missing color-related parameters are shown as -9.
\end{table*}

The published catalog of most variable CRF objects includes robust statistics of variability derived for 5033 sources. Table \ref{para.tab} provides the leading portion of this catalog for convenient reference. The entire catalog is available online. For each source, the Gaia source identifier and the mean RA and Decl. coordinates in degrees are copied from the Gaia CRF catalog. The derived parameters are: the median $\hat G$ magnitude, its robust standard deviation smad($\hat G$) (Eq. \ref{hatG.eq}), the score of variability in $G$ ($\hat Z$, Eq. \ref{z.eq}), the number of accepted observations in $G$, the median BP$-$RP color ($\hat C$, Eq. \ref{c.eq}) and its robust standard deviation, separate robust standard deviations of the measured BP and RP magnitudes, and the Spearman's rho correlations $\rho(G$,BP$-$RP) and $\rho$(BP,RP). A small fraction of color-related values are not abailable, in which case they are replaced by $-9$. 

Investigation of some unusual light curves and peculiar photometric properties revealed that some of the listed variable sources are in fact stellar contaminants. The source Gaia 5050179952895664000 shows an atypical sinusoidal light curve component superposed with a longer-term trend. It matches a SIMBAD object DDB2002, which is a known carbon star in the Fornax dwarf galaxy. The only source in the catalog with a negative correlation $\rho$(BP,RP) is 4657348637547570560, an extreme infrared source with a median BP$-$RP$=5.38$ mag. This source matches IRAS 05558$-$7000, a superluminous AGB star in the Large Magellanic Cloud. The third reddest source is 542175166549664512, which is probably a Galactic young stellar object (YSO), not a quasar. The CRF source 4111272369389570816 is another exceptionally red object (BP$-$RP=4.06 mag), which shows a periodic component in the light curve \citep{2023A&A...674A..15L}, and matches a Mira-type star in the OGLE collection of long-period variables \citep{2022ApJS..260...46I}.

\section{Comparative analysis of the ICRF3 sample}

The intersection of Gaia CRF3 with the ICRF3 catalog is of special interest for the task of improvement and maintenance of the fundamental reference frame. These relatively rare sources, which are both optically visible and radio-loud, are used to align the Gaia coordinate frame with that of the ICRS. Their photometric and astrometric properties may be intertwined, as quasars and AGNs are generally neither point-like sources nor intrinsically stable structures \citep{2012MmSAI..83..952M}. To better understand the complex astrometric effects of ICRF3 quasars, we need to map the photometric parameter space and correlate it with detailed spectroscopic and astrometric data. 

Using the IERS name cross-identification in the Gaia CRF3 catalog, we find 2690 common objects. This is only 0.32\% of the general sample. The selected MVO-sample (5033 sources) is 0.60\% of the general CRF-sample. We would expect to find 16 ICRF3 objects in the MVO catalog if they were randomly chosen. Instead, we find 488 common ICRF3/MVO objects. This would seem to indicate that the ICRF3 optical counterparts tend to be bright and highly variable compared to the general CRF-sample. A closer look at the properties of the 2690 sources reveals a more complicated picture.

\begin{figure*}
\includegraphics[width=0.45\linewidth]{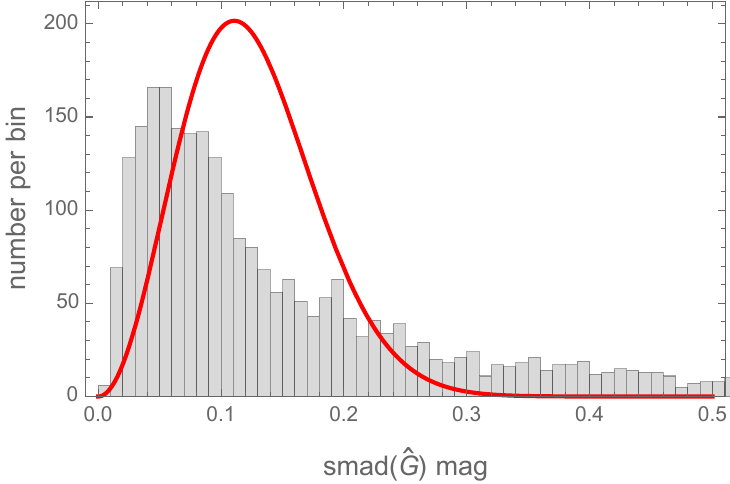}
\includegraphics[width=0.45\linewidth]{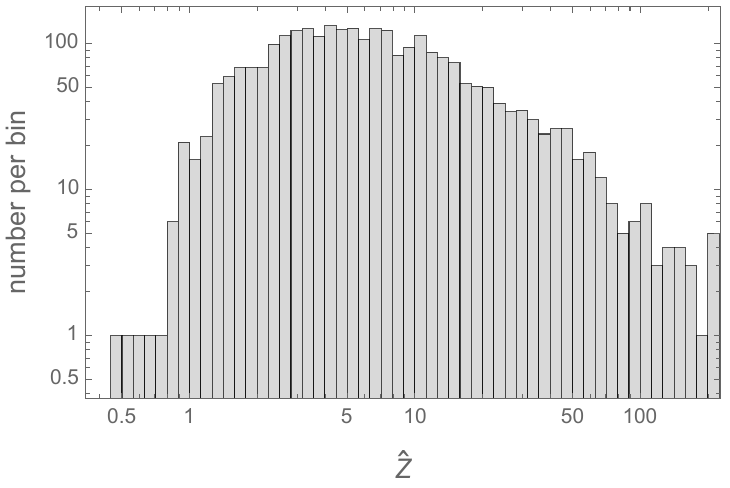}
\caption{Histograms of robust variability parameters in $G$ for the ICRF3 subset of Gaia CRF3. Left panel: dispersion of measured magnitudes smad$(\hat G)$. The red curve is copied and re-normalized from Fig. \ref{G.fig} to visual comparison with the general sample. Right panel: normalized dispersion of measured magnitudes $\hat Z$.}
\label{ICRF.fig}
\end{figure*}

Fig. \ref{ICRF.fig} displays the histograms of two variability parameters, the variability amplitude smad($\hat G$) and the normalized score $\hat Z$ in the same axes and bin widths as the general sample counterparts in Fig. \ref{G.fig}. For the left panel, the same best-fitting analytical distribution function is reproduced with re-normalization to assist the comparison. The peak of the smad($\hat G$) distribution has moved to smaller values for the ICRF3 sample, meaning that a significant fraction of these sources have small variability amplitudes. This change is countered by a massive tail stretching far beyond 0.2 mag. The median smad($\hat G$) has barely changed between the samples (0.105 mag for CRF versus 0.110 mag for ICRF), but the outer quantiles are much higher for ICRF, with the 0.75 value being 0.230 mag against 0.145 mag for CRF. Even more dramatic changes are seen in the distributions of $\hat Z$, which quantifies the statistical significance (similar to signal-to-noise ratio) of the absolute variability amplitudes. Here, the median value changes from 2.7 for CRF to 5.6 for ICRF. This difference is mostly due to the relative median magnitudes of the ICRF counterparts, which are almost 2 mag brighter. Thus, the excessive rate of highly variable sources among the ICRF counterparts has dual origin: roughly a quarter of these sources are indeed exceptionally variable, and many have brighter magnitudes and smaller formal errors.

\section{Optical variability versus redshift}

This part of analysis is limited to the brighter one-third of the general CRF-sample. We find 0.307 million objects in the general CRF-sample with median $\hat G$ magnitudes brighter than 19.6. The distribution of BP$-$RP colors is peaked at 0.6 mag, but a significant subset is redder than 1 mag, and the increasingly uncertain measurement in BP at the faint end may bias the variability statistics by, for example, the survival selection. The source of redshifts is the Quaia catalog of 1.296 million quasars \citep{2024ApJ...964...69S}, which is based on the rough spectroscopic estimates from the Gaia QSO candidates \citep{2023A&A...674A..41G} supplemented with mid infrared photometry from unWISE \citep{2014AJ....147..108L, 2019PASP..131l4504M}. The intersection of Quaia with the bright part of CRF, which is here called the $z$-sample, includes 0.298 million objects. 

Possible systematic dependencies of variability parameters on redshift are investigated using robust quantiles of the empirical distributions, which are obviously non-Gaussian. The entire $z$-sample if sorted by increasing redshift $z$ (taken from Quaia). The sorted list is partitioned into 150 bins of equal size. For each bin, the distribution of a parameter of interest is evaluated by computing the \{0.158655, 0.5, 0.841345\}
quantiles. The 0.5-quantile is by definition equivalent to the median value, which is a robust analog of the sample mean. The bracketing quantiles are chosen to correspond to the $\pm 1\,\sigma$ interval of the normal distribution. These can be considered as analogs of the statistical dispersion around the median.

\begin{figure*}
\includegraphics[width=0.31\linewidth]{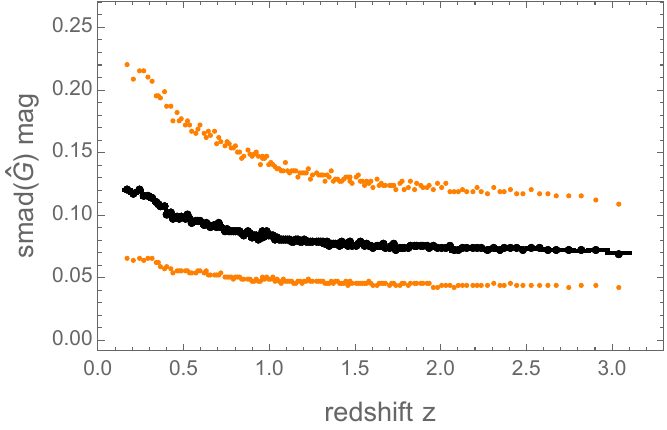}
\includegraphics[width=0.31\linewidth]{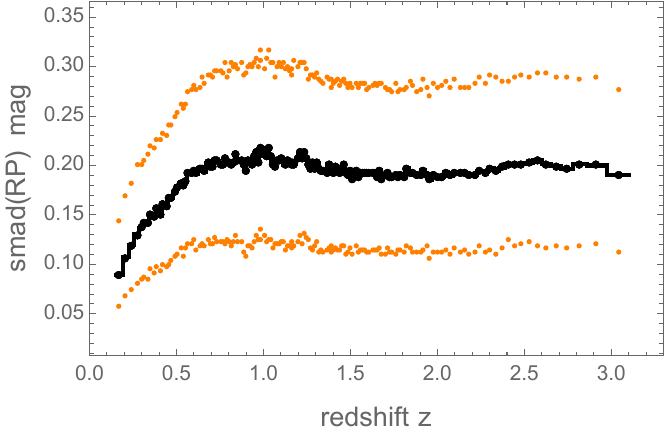}
\includegraphics[width=0.31\linewidth]{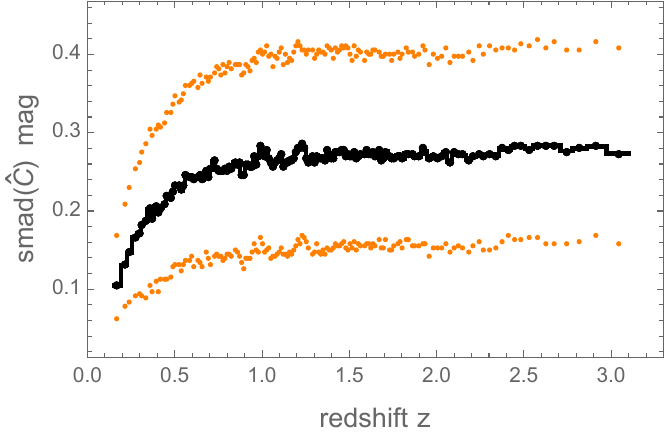}
\caption{Robust photometric and variability parameters versus redshift for 0.298 million Gaia CRF quasars. The black dots connected with step lines are the median values for each bin of the sample sorted by $z$. The orange dots show the 0.16--0.84 quantile dispersion limits corresponding to $\pm\,1\sigma$ interval. Left: variability amplitude in $G$ magnitude. Middle: variability amplitude in \rp\ magnitude. Right: variability amplitude in \bp$-$\rp\ color.}
\label{z1.fig}
\end{figure*}

\begin{figure*}
\includegraphics[width=0.31\linewidth]{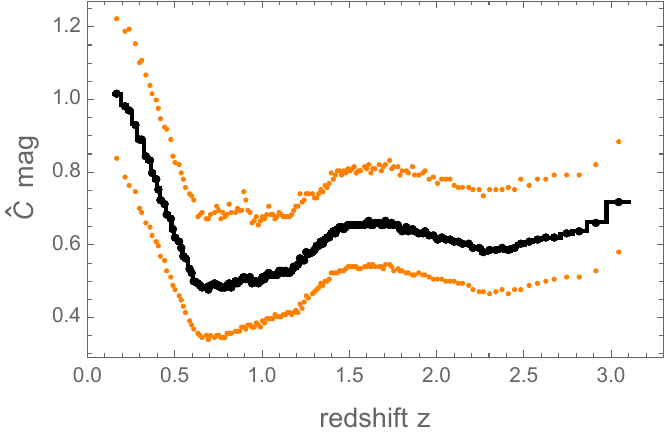}
\includegraphics[width=0.31\linewidth]{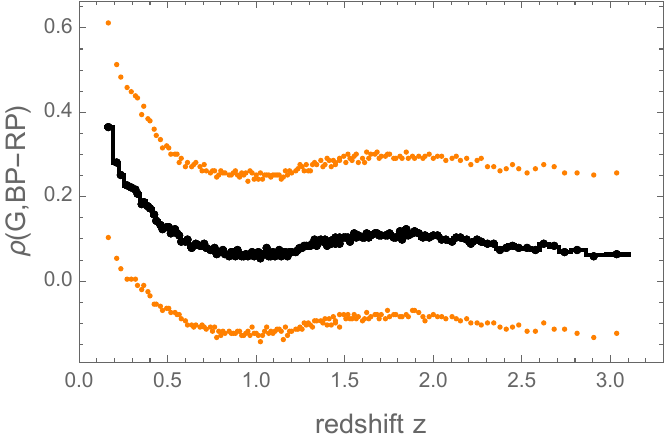}
\includegraphics[width=0.31\linewidth]{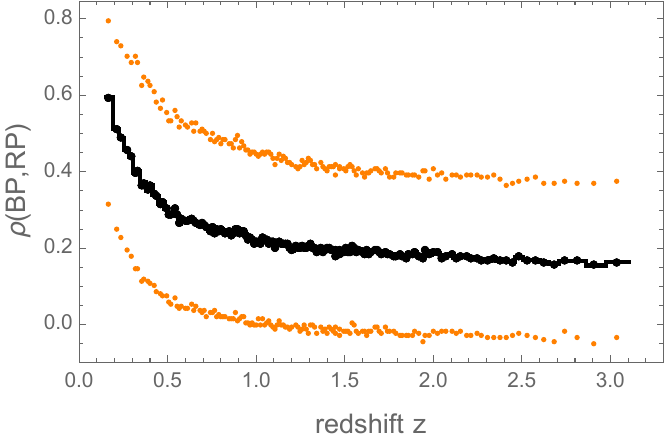}
\caption{Robust photometric and variability parameters versus redshift for 0.298 million Gaia CRF quasars. The black dots connected with step lines are the median values for each bin of the sample sorted by $z$. The orange dots show the 0.16--0.84 quantile dispersion limits corresponding to $\pm\,1\sigma$ interval. Left: median \bp$-$\rp\ color. Middle: Spearman's correlation $\rho$($G$,BP$-$RP). Right: Spearman's correlation $\rho$(BP,RP).}
\label{z2.fig}
\end{figure*}

Some of the results are shown in Figs. \ref{z1.fig} and \ref{z2.fig}. While the broadband variability amplitude smad($\hat G$) is a monotonic declining function of $z$, which is in agreement with the results in \citep{2021AJ....162...21B}, a very different behavior is seen for the red magnitude \rp\ and the color
\bp$-$\rp. The latter are fairly flat for $z>1$, but steeply increase for the closer sources. This may seem puzzling given that the BP and RP bands largely overlap with the $G$ band. The most likely explanation is an instrumental effect unrelated to the physics of quasars. The aperture-type Gaia photometry in BP and RP is perturbed and biased by the resolved images of host galaxies, which is reflected in greatly elevated {\tt phot\_bp\_rp\_excess\_factor} values. 

The sagging dependence of smad(RP) at $z$ between 1.2 and 2.2 may be genuine, however. It corresponds to bumps in the relations of color $\hat C$ and $\rho(G$,BP$-$RP). A single dominant mechanism of emission in the accretion disk, combined with the redshifted bandpass interval in the quasar rest frame can give the observed behavior. The color-redshift relation for quasars have been investigated in the literature based on SDSS photometry \citep{2004ChJAA...4...17W, 2004ApJS..155..243W}. Our result shown in Fig. \ref{z2.fig}, left, is generally consistent with the previously detected relation for the SDSS $r'-i'$ color. This behavior is explained by various spectral lines entering and leaving the photometric band at specific redshifts. The observed dependence of $\rho$(BP,RP) on $z$ is monotonical and featureless. A more complex behavior is seen for $\rho(G,$BP$-$RP), which has a local minimum at $z\simeq 0.9$ and a local maximum at $z\simeq 1.8$. This undulation is likely related to the intrinsic color variability of quasars \citep{2012ApJ...744..147S, 2014ApJ...792...54S}. The underlying mechanism is the impact of various spectral features that are present within the photometric bandpass, combined with the generally bluer continuum when the source becomes brighter. This explains the conspicuous anti-correlation between the median dependencies for smad(RP) and  $\rho(G,$BP$-$RP). Compared to the analogous Fig. 2 in \citep{2012ApJ...744..147S}, the median $\rho(G,$BP$-$RP) is flatter, which may be caused by the broader photometric bands of Gaia.

\section{Results and Conclusions} \label{con.sec}

The presence of a large population of CRF quasars and AGN with modest degrees of optical variability is a surprising result with practical implications for the development and maintenance of the Fundamental reference frame. A dichotomy of quasars and AGNs with regard to variability degree can be suggested, although we do not find a clear separation of these populations in any of the tested parameters. Some of the quasars are indeed weakly variable sources in the optical. For example, the object in the ICRF-sample with the smallest smad($\hat G$) of 0.007 mag is the bright relatively nearby BL Lac source QSO B2128$-$123. Its remarkable constancy has been known from ground-based observations \citep{1985AJ.....90...39M}, which also revealed that the amplitude of variability is greatly higher in the reddest Johnson band $I$. 

\begin{figure*}
\includegraphics[width=0.45\linewidth]{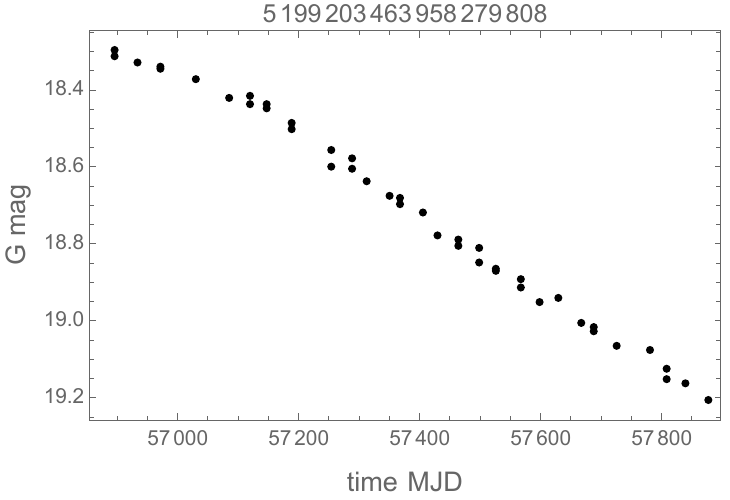}
\includegraphics[width=0.45\linewidth]{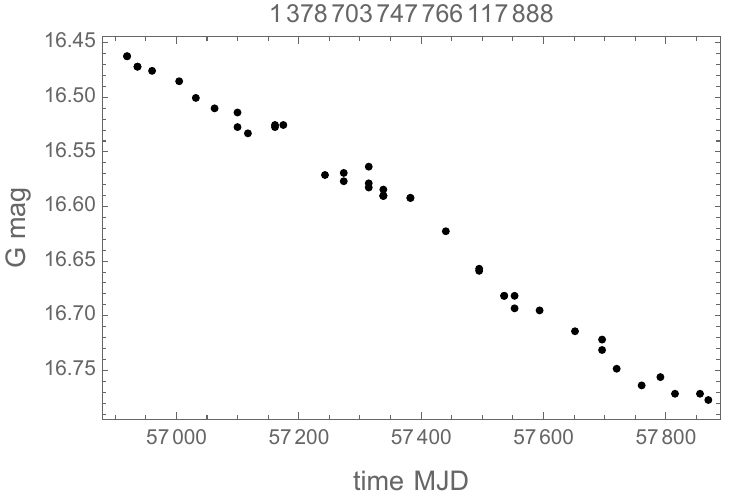}
\caption{Gaia DR3 light curves in $G$ magnitude of two sources from the MVO-sample with the highest Spearman's correlation value 0.9996.}
\label{spmax.fig}
\end{figure*}

There are identified caveats to our results. The general variability amplitudes may be underestimated for the CRF-sample. The presence of marginally faint objects is a source of estimation bias, especially when the BP-band is concerned. The detection threshold of 1 $e^-$ s$^{-1}$ results in a lopsided selection of the brighter flux measures, thus effectively shifting the mean and median values \citep{2021A&A...649A...3R, 2021A&A...649A...5F}. The brightness minima are not properly captured in the accepted detections. Also, the time span of Gaia DR3 epoch photometry is less than 3 yr, which causes the variability magnitudes to be systematically underestimated. The characteristic time of quasar variations may be longer than the time series window, and what is available in Gaia light curves is only a section of the full range. This is confirmed by a special calculation on the MVO-sample of the 5033 most variable sources. The Spearman's correlation coefficient of the observed light curves in $G$ magnitude was computed for each of the 12.66 million possible pairs of sources. This coefficient quantifies the degree of coherence, or sequential homomorphism, of the given pair's light curves. The histogram of thus computed correlations is non-uniform in the $[-1,+1]$ support interval showing two prominent symmetric peaks at $\pm 0.8$. Hence, a large number of unrelated sources show light curves that look similar, or alike. This population is composed of quasars with longer time scales of variation, which are above the time coverage of DR3. Fig. \ref{spmax.fig} shows the light curves of the pair with the largest positive correlation of 0.9996. Obviously, the morphological similarity is caused by the almost linear decline of their brightness during the $\sim 1000$ days of observation. The full amplitude of variation is likely to be much higher. Consequently, the commonly used first-order structure functions would give misleading results for sources like the ones shown in Fig. \ref{spmax.fig}, because they do not remove linear or polynomial trends \citep{1985ApJ...296...46S}. Either higher-order structure functions or more sophisticated statistical analysis than the underlying Allan variance \citep{1966IEEEP..54..221A} should be employed.

The currently available Gaia DR3 light curves represent a small fraction of the photometric data that will be published at the end of this mission. Longer time coverage and larger data samples are needed to employ more elaborate methods of statistical analysis based on the nonparametric autocorrelation functions, including sequence differencing, ARMA and ARIMA random process models \citep{2018FrP.....6...80F}. The irregular cadence of Gaia measurements will still represent a significant limitation, and the techniques based on the continuous random process models with homoscedastic or heteroscedastic options (e.g., CARMA and GARCH) may prove more efficient. Machine learning methods can be explored to quantitatively classify the quasar light curves into major types, such as bursting, noisy, and long-term coherent. Reaping the advantages of space-based photometry, inroads can be made into the quasi-periodic behavior of some reference quasars, which has been found in precision VLBI astrometry \citep{2024PASP..136e4503M}.

The majority of Gaia CRF sample show positive correlation between the $G$ magnitude and BP$-$RP color. This is consistent with the previously discussed ``brighter-bluer" effect (BWB) seen in ground-based photometric surveys including much smaller samples of quasars. However, we find a significant fraction of sources that are not compliant with this rule, i.e., they statistically become redder when they are brighter (RWB). For the brighter one-third of the $BR$-sample including 0.307 million objects, where the photometric survival bias is much reduced, 30\% of the sources have negative $\rho(G,$BP$-$RP) correlation values. This result is in tension with the rate of 6\% for RWB objects estimated by \citet{2016ApJ...822...26G} from a selection of quasars with SDSS epoch photometry and spectra. We note that the sample investigated in this paper is a hundred times larger; on the other hand, hidden calibration issues with the BP, RP data cannot be precluded.
The rate of RWB sources is even higher (39\%) in the ICRF-sample. Alternatively, contributions from two different physical mechanisms of optical variability (thermal disk instability versus non-thermal emission from the jet) may be involved \citep{2011A&A...534A..59G, 2013A&A...554A..51G}. It is possible that the BL Lacertae objects are in majority relative to the FSRQ objects in the CRF-sample (specifically, 60--70\%) given the distribution of magnitude--color correlation \citep{2022MNRAS.510.1791N}. \citet{2017ApJS..229...21X} determined from a decade-long photometry of the optically brightest quasar 3C 273\footnote{This source is not present in our analysis because almost all its measurements are flagged out as ``noisy" in Gaia DR3.} that the degree of the BWB effect is inversely correlated with the characteristic time scale. The strongest positive correlation is seen for intra-night variations, while the effect disappears on the scale of years. With the total duration of 1000 d and time resolution about one month, the Gaia epoch photometry is probing the longest characteristic times, where the BWB may be much diluted.

A stronger positive correlation is found between the variations in BP and RP bands, indicating mostly coherent changes in the two parts of the optical spectrum. This is consistent with the standard model of accretion disk around the central black hole \citep{1973A&A....24..337S}, in which the optical continuum is caused by the thermal radiation of the accreting material and outflows. A finite fraction  of the CRF sample, however, seems to buck this trend with negative $\rho$(BP,RP) values even after a faint magnitude cut, which is expected to remove the known systematic errors in Gaia. This population of discordant sources deserves further investigation. They mainly occur among the objects with smad($\hat G)\simeq 0.06$ and high redshifts. Almost all of them are fainter than $\hat G=19$ mag, so the survival bias of \bp\ magnitudes may still be responsible.

Our comparative analysis of the ICRF3 subset of Gaia CRF confirms the main conclusions of \citep{2021AJ....162...21B} based on an elaborate processing of the Pan-STARRS multi-band epoch photometry. The modal values of variability amplitudes (in terms of rms or standard deviation) are indeed below 0.1 mag, so that many ICRF counterparts are weakly variable sources. However, a dichotomy may be present in this sample, with a quarter of sources showing extreme degrees of variability. Further studies are needed to shed light on possible physical origins of this dichotomy. One obvious possibility is that the low-variability ICRF3 radio sources are mostly AGNs residing in nearby galaxies, there the lion's share of the optical flux comes from the constant host galaxy. This can be tested by correlating the variability amplitude with the redshift. In confirmation of the main conclusion by \citet{2021AJ....162...21B}, we find that variability amplitude is a steadily declining function of redshift. Therefore, either quasars were less variable in the early universe, or the far-UV portion of the redshifted quasar emission is intrinsically more constant. Alternatively, the cosmological time dilation \citep{2023NatAs...7.1265L} can provide an elegant explanation, given the limited time span of the available Gaia light curves. Assuming that all quasars have a characteristic time variability distribution, which is independent of the distance, an increasing fraction of sources' light curves become unresolved in time within the span of 1000 days with increasing redshift.

\section{Acknowledgments}
This work has made use of data from the European Space Agency (ESA) mission {\it Gaia} (\url{https://www.cosmos.esa.int/gaia}), processed by the {\it Gaia} Data Processing and Analysis Consortium (DPAC, \url{https://www.cosmos.esa.int/web/gaia/dpac/consortium}). Funding for the DPAC has been provided by national institutions, in particular the institutions participating in the {\it Gaia} Multilateral Agreement. This research has made use of the {\it VizieR} catalogue access tool, CDS, Strasbourg, France. This work supports USNO's ongoing research into the celestial reference frame and geodesy.


\bibliography{main}
\bibliographystyle{aasjournal}

\newpage

\end{document}